\documentclass[acmsmall]{acmart}
\usepackage{booktabs}
\usepackage{xspace}

\usepackage{multirow}
\usepackage{enumitem}
\usepackage[ruled,vlined,linesnumbered,lined, commentsnumbered]{algorithm2e}
\usepackage[noend]{algpseudocode}
\usepackage{bm}
\usepackage{subfigure}

\AtBeginDocument{%
  \providecommand\BibTeX{{%
    \normalfont B\kern-0.5em{\scshape i\kern-0.25em b}\kern-0.8em\TeX}}}

\setcopyright{acmcopyright}
\copyrightyear{2018}
\acmYear{2018}
\acmDOI{XXXXXXX.XXXXXXX}

\newcommand{\proposed}{\textsf{PerFedRec++}\xspace}

\begin{document}

\title{PerFedRec++: Enhancing Personalized Federated Recommendation with Self-Supervised Pre-Training}

\author{Sichun Luo}
\affiliation{
  \institution{City University of Hong Kong}
  \city{Hong Kong}
  \state{Kowloon}
  \country{China}
  \postcode{999077}
}
\email{sichun.luo@my.cityu.edu.hk}

\author{Yuanzhang Xiao}
\affiliation{%
\department{Hawaii Advanced Wireless Technologies Institute}
  \institution{University of Hawaii at Manoa}
  \city{Honolulu}
  \state{HI}
  \country{USA}
}
\email{yxiao8@hawaii.edu}

\author{Xinyi Zhang}
\affiliation{%
\department{Department of Accounting}
  \institution{Capital University of Economics and Business}
  \city{Beijing}
  \country{China}
}
\email{zhangxinyi@cueb.edu.cn}

\author{Yang Liu}
\affiliation{
  \institution{ Tsinghua University }
  \city{ Beijing }
  \state{ Beijing }
  \country{China}
  \postcode{999077}
}
\email{liuyang.princeton@gmail.com}

\author{Wenbo Ding}
\affiliation{
  \institution{ Tsinghua University }
  \city{ Shenzhen }
  \state{ Guangdong }
  \country{China}
  \postcode{518071}
}
\email{ding.wenbo@sz.tsinghua.edu.cn}


\author{Linqi Song}
\authornote{Corresponding author.}
\affiliation{
  \institution{City University of Hong Kong}
  \city{Hong Kong}
  \state{Kowloon}
  \country{China}
  \postcode{999077}
}
\email{linqisong@cityu.edu.hk}

\renewcommand{\shortauthors}{Luo et al.}

\begin{abstract}
Federated recommendation systems employ federated learning techniques to safeguard user privacy by transmitting model parameters instead of raw user data between user devices and the central server. Nevertheless, the current federated recommender system faces three significant challenges: (1) \textit{heterogeneity and personalization}: the heterogeneity of users' attributes and local data necessitates the acquisition of personalized models to improve the performance of federated recommendation; (2) \textit{model performance degradation}: the privacy-preserving protocol design in the federated recommendation, such as pseudo item labeling and differential privacy, would deteriorate the model performance; (3) \textit{communication bottleneck}: the standard federated recommendation algorithm can have a high communication overhead. Previous studies have attempted to address these issues, but none have been able to solve them simultaneously.

In this paper, we propose a novel framework, named \proposed, to enhance the personalized federated recommendation with self-supervised pre-training. Specifically, we utilize the privacy-preserving mechanism of federated recommender systems to generate two augmented graph views, which are used as contrastive tasks in self-supervised graph learning to pre-train the model. Pre-training enhances the performance of federated models by improving the uniformity of representation learning.
Also, by providing a better initial state for federated training, pre-training makes the overall training converge faster, thus alleviating the heavy communication burden. We then construct a collaborative graph to learn the client representation through a federated graph neural network. Based on these learned representations, we cluster users into different user groups and learn personalized models for each cluster. Each user learns a personalized model by combining the global federated model, the cluster-level federated model, and its own fine-tuned local model. Experiments on three real-world datasets show that our proposed method achieves superior performance over existing methods.

\end{abstract}



\authorsaddresses{
Authors' addresses: 
S. Luo,
Department of Computer Science, City University of Hong Kong, Hong Kong, China; email: sichun.luo@my.city.edu.hk;
Y. Xiao,
Hawaii Advanced Wireless Technologies Institute, University of Hawaii, Honolulu, HI, USA,
email: yxiao8@hawaii.edu; 
X. Zhang,
Department of Accounting, Capital University of Economics and Business,
Beijing, China, email:
zhangxinyi@cueb.edu.cn;
Y. Liu,
Institute for AI Industry Research,
Tsinghua University,
Beijing, China, email:
liuyang.princeton@gmail.com;
W. Ding,
Institute of Data and Information, Tsinghua Shenzhen International Graduate School, China, email: ding.wenbo@sz.tsinghua.edu.cn;
L. Song,
Department of Computer Science,
City University of Hong Kong,
Hong Kong, China, email:
linqi.song@cityu.edu.hk.
}




\maketitle

\section{Introduction}

Recommender systems aim to provide personalized recommendations to users based on their preferences and past behaviors. These systems have become ubiquitous in e-commerce, social media, and other online platforms, with the potential to enhance user experience and increase revenue \cite{schafer1999recommender, ma2018point,luo2022hysage}.
However, the collection and use of user data by conventional recommender systems have raised serious concerns about user privacy. In response to these concerns, regulations such as the General Data Protection Regulation (GDPR)\footnote{https://gdpr-info.eu/} and the California Consumer Privacy Act (CCPA)\footnote{https://oag.ca.gov/privacy/ccpa} have been enacted to protect user privacy. 

To address the privacy issues, Federated Learning (FL) has emerged as a promising privacy-preserving paradigm that differs from conventional machine learning techniques \cite{mcmahan2017communication}. FL techniques localize data and transmit only the model updates to a central server, thereby ensuring the privacy of user data. The integration of FL with recommender systems has given rise to a nascent research area, known as federated recommendation \cite{yang2020federated}. Federated recommendation is a novel approach that aims to improve the filtering of useful information by users while preserving their privacy. This technique follows the principles of federated learning, where recommendation models, rather than raw data, are exchanged between user devices and a central server. This paradigm has found diverse applications in various areas, such as content recommendations \cite{tan2020federated,ali2021federated}, mobile crowdsourcing task recommendations \cite{zhang2020pfcrowd,guo2020fedcrowd}, and autonomous driving strategy recommendations \cite{savazzi2021opportunities}.
Additionally, various classical recommendation algorithms have been extended to the federated setting, such as federated collaborative filtering \cite{ammad2019federated,minto2021stronger}, federated matrix factorization \cite{chai2020secure,li2021federated,du2021federated}, and federated graph neural network (GNN) \cite{wu2021fedgnn}. 
However, current federated recommendation systems have several limitations that hinder their effectiveness:
\begin{itemize}
\item
\textbf{Heterogeneity and Personalization:} The non-iid distribution of local data due to each client's unique item interests presents a significant challenge in developing personalized federated learning processes \cite{yang2019federated,fallah2020personalized}. Addressing this heterogeneity and personalization problem requires developing models that can effectively capture individual clients' preferences while also considering the overall objective regarding all the clients' preferences.
\item \textbf{Model Performance Degradation:} Federated recommender systems typically employ standard GNN structures which may not be capable of learning uniform representations, while previous works show that model performance improves when the embeddings are more uniformly distributed in the latent space \cite{yu2022xsimgcl,yu2022graph2,luo2022personalized}.
Additionally, the privacy-preserving mechanisms in federated recommender systems such as differential privacy and pseudo item labeling can also lead to deterioration in the model performance~\cite{wu2021fedgnn}.
\item
\textbf{High Communication Costs:} In the realm of federated recommender systems, most of them randomly select users for federated learning from all available users, resulting in high communication costs. This approach is not communication-efficient and can result in substantial delays in the model training process.
\end{itemize}
Overcoming these limitations is crucial for the development of effective federated recommendation systems. These challenges highlight the need for novel algorithms and architectures that can address heterogeneity and personalization, learn uniform representations for effective federated recommendation, and improve communication efficiency.






Self-supervised learning is a rapidly advancing field that has gained significant attention in recent years \cite{liu2021self,jaiswal2020survey}. Unlike traditional supervised learning, where labeled data is required for training, self-supervised learning leverages unlabeled data to learn representations that can be used for downstream tasks \cite{jing2020self,misra2020self}. Graph contrastive learning is a promising self-supervised learning technique that aims to learn low-dimensional representations of graphs by maximizing the agreement between similar pairs of nodes and minimizing the agreement between dissimilar pairs of nodes \cite{xie2022self}. In graph contrastive learning, a graph encoder is used to map each node in the graph to a low-dimensional embedding, and a contrastive loss function is utilized to optimize the agreement between the embeddings of similar nodes (\textit{i.e.}, positive samples) and the disagreement between the embeddings of dissimilar nodes (\textit{i.e.}, negative samples).  Unlike other graph representation learning techniques, graph contrastive learning does not require labeled data and can effectively capture the underlying structural properties of the graph. As such, graph contrastive learning has been successfully applied to various graph-related tasks, including graph clustering \cite{zhong2021graph}, node classification \cite{you2020graph}, and recommendation \cite{wu2021self}. Nevertheless, applying graph contrastive learning to federated recommendation systems is a non-trivial problem. In the standard federated recommendation setting, clients are unable to access the latent embeddings of other clients, thereby making it hard to construct negative samples.
To solve this problem, we propose a self-supervised pre-training stage in which the server conducts self-supervised pre-training using the embeddings uploaded by the clients. In this way, the server would have sufficient negative samples to perform graph contrastive learning.


In this paper, we propose \proposed, a novel framework designed for improving personalized federated recommendations via self-supervised pre-training. 
Our key innovation is to utilize the byproduct of the privacy-preserving mechanism in federated recommender systems as data augmentation. Specifically, the privacy-preserving mechanism, such as client selection, pseudo item labeling, and differential privacy, can generate different ``views'' of the system. This is analogous to node dropout, edge perturbation, and noise injection in data augmentation. Then we employ a contrastive loss to maximize the agreement between different views of the same node, as compared to the other nodes. 
By incorporating self-supervised pre-training, \proposed are able to learn more uniform representations and thus improve the model performance.
Meanwhile, self-supervised pre-training could provide a better initial state for later federated training, thus could make the model converge faster and thus alleviating the communication burden.
Moreover, \proposed utilizes collaborative information, \textit{i.e.}, user-item interactions, to learn user representations through a local GNN. These learned representations are then used to group similar users into clusters, each of which is assigned a cluster-level federated recommendation model, while the server obtains a global model. Finally, each user combines their local model, the cluster-level model, and the global model to obtain a personalized recommendation.




 
In a nutshell, the key contributions of this paper are summarized as follows:\footnote{Part of this paper has been presented in ACM CIKM 2022 \cite{luo2022personalized}.}

\begin{itemize}
\item
We propose \proposed, a framework that incorporates self-supervised pre-training into the personalized federated recommendation task to learn more uniform representations and improve model performance. 
\item \proposed introduces three ad-hoc data augmentation techniques to generate augmented graph views, and adopts the contrastive loss to mine contrastive signals from the input data. The pre-training process could enhance the model performance and ease the communication burden, which could address the challenge of model performance degradation and high communication costs.
\item
To address the challenge of heterogeneity, our proposed framework employs user-item interactions to learn user representations, group similar users into clusters, and assign cluster-level and global recommendation models to obtain personalized recommendations.
\item
We conduct extensive experiments on real-world datasets to validate the  effectiveness and efficiency of \proposed. Experimental results from three real-world datasets demonstrate the performance improvement over a variety of existing methods.
\end{itemize}

\section{Related Work}

\subsection{GNN based Recommendations}

Graph neural networks have emerged as powerful architectures for modeling recommendation data, and are supplanting multi-layer perceptron (MLP)-based models in academia \cite{wu2022graph}. GNNs have elevated neural recommender systems to a new level, and a large number of recommendation models that adopt GNNs claim to have achieved state-of-the-art performance in various subfields. 
GCMC \cite{berg2017graph} formulated the recommendation task from the point of view of link prediction on graphs, and proposed a graph-based auto-encoder framework for
matrix completion.
Wang \textit{et al.} proposed a novel graph-based collaborative filtering approach that incorporates user-item interactions and graph structure information using graph neural networks \cite{wang2019neural}.
LightGCN \cite{he2020lightgcn} removes redundant operations such as transformation matrices and nonlinear activation functions, which has proven to be efficient and effective. It is the most popular method due to its simplicity and decent performance.
Despite the differences in implementation, these GNN-based recommendation models share the common idea of acquiring information from neighbors in the user-item graph layer by layer to refine target node embeddings and achieve graph reasoning.

\subsection{Self-Supervised Learning for Recommender Systems}

Contrastive learning is a type of self-supervised learning that aims to learn informative representations by contrasting positive and negative examples \cite{jaiswal2020survey}. The standard contrastive learning paradigm generates two augmented views from the input data, and then maximizes the agreement between these views. This approach has exhibited significant success in a range of computer vision and natural language processing tasks by leveraging the idea of maximizing the similarity between instances from the same class and minimizing the similarity between instances from different classes \cite{chen2020simple, chen2020improved, giorgi2020declutr}. Contrastive learning has also been integrated into recommender systems to improve recommendation performance. For instance, CL4SRec \cite{xie2022contrastive} incorporates self-supervision signals to alleviate the data sparsity issue and enhance model performance in sequential recommendation.

Recently, contrastive learning has been extended to GNN based recommender systems. Wu \textit{et al.} \cite{wu2021self} proposed a self-supervised graph learning (SGL) model that utilizes graph neural networks to model user-item interactions and employs graph contrastive learning to optimize the model parameters. SimGCL \cite{yu2022graph2} introduces a different data augmentation approach, \textit{i.e.}, injecting random noises into embeddings as an augmentation-free method for graph contrastive learning. xSimGCL \cite{yu2022xsimgcl} further simplifies and enhances SimGCL by only augmenting on one view. These studies have demonstrated that graph contrastive learning is an effective technique for improving the accuracy and efficiency of recommendation systems.

Overall, these recent papers demonstrate the effectiveness of using self-supervised learning for recommendation, particularly in the context of modeling user-item interactions with graph neural networks. These techniques have the potential to improve the accuracy and effectiveness of recommendation systems by better capturing complex relationships between users and items.
However, applying contrastive learning in the recommendation scenario is not straightforward, since there is no uniform data augmentation method for contrastive learning under all situations, and thus needs to design by the problem situation case by case.
To address this issue, we propose a tailor-made data augmentation method for the federated recommendation that takes into account the unique characteristics of federated recommender systems.

\subsection{Federated  Recommender Systems}
Federated recommendation is a rapidly growing research area that aims to enhance recommendation systems by leveraging the power of federated learning \cite{sun2022survey, luo2022towards}. Two pioneering works investigating a novel federated learning framework to learn the user/item embeddings for a recommender system are Federated Collaborative Filtering \cite{ammad2019federated} and Federated Matrix Factorization \cite{chai2020secure}. Both works develop federated learning on top of factorization of the user-item rating matrix. To achieve federated learning, they propose that users' ratings should be stored locally, with user embeddings trained locally, and the server only retaining the item embeddings. This training framework protects user privacy as there is no transfer of users' interactions. Differential privacy \cite{dwork2006differential} is also incorporated to enhance the privacy-preserving ability.

Recent advancements in federated learning for recommendation have also explored the use of graph neural networks to model user-item interactions. Wu \textit{et al.} proposed a Federated Graph Neural Network (FedGNN) that leverages GNNs and federated learning to enhance the accuracy and privacy of the federated recommendation system \cite{wu2021fedgnn}.
Moreover, there have been efforts to address the challenges of federated learning for recommendation, such as data heterogeneity and communication efficiency. For instance, PerFedRec \cite{luo2022personalized} proposed techniques to deal with non-IID (non-identically distributed) data in a federated learning setting, while FedFast \cite{muhammad2020fedfast} explored the use of communication-efficient algorithms to reduce the communication overhead in federated learning.
Table~\ref{tab:model_compare} provides a comprehensive comparison between a selection of prominent recommender systems and federated learning approaches.

Despite the considerable advancements in federated recommender systems, their potential remains constrained insofar as they struggle to effectively integrate personalization, efficiency, and accuracy into a cohesive framework. Consequently, our research endeavors to develop a novel model that can effectively reconcile these limitations simultaneously.

\begin{table}[htbp]
\caption{Comparison of representative models with respect to GNN, self-supervised learning, and federated setting.}\label{tab:model_compare}
  \centering
\begin{tabular}{c|cccc}
   \toprule 
    & GNN & Self-Supervised Learning  & Federated Setting  \\
   \midrule
   LightGCN \cite{he2020lightgcn}& \checkmark & $\times$  & $\times$  \\
   SGL \cite{wu2021self}& \checkmark & \checkmark & $\times$  \\
   SimGCL \cite{yu2022graph2}& \checkmark & \checkmark  & $\times$  \\
xSimGCL \cite{yu2022xsimgcl}& \checkmark & \checkmark &  $\times$  \\
   FCF \cite{ammad2019federated}& $\times$ & $\times$ &  \checkmark   \\
   FedMF \cite{chai2020secure}& $\times$ & $\times$ &  \checkmark  \\

   FedGNN \cite{wu2021fedgnn}& \checkmark & $\times$ &  \checkmark \\
   FedCL \cite{wu2022fedcl}& $\times$ & \checkmark &  \checkmark \\
   
   PerFedRec \cite{luo2022personalized}& \checkmark & $\times$ & \checkmark \\
   \proposed & \checkmark & \checkmark &  \checkmark \\

   \bottomrule
\end{tabular}
\end{table}

\section{ Preliminaries }


We define $\mathcal{U} = \{1, \ldots, N\}$ and $\mathcal{I} = \{1, \ldots, M\}$ as the sets of users and items, respectively. Each user $n \in \mathcal{U}$ has some historical interactions with items (\textit{e.g.}, rating the items). The historical interactions is represented by the matrix $\mathbf{R} \in \{0,1\}^{N \times M}$, where each entry $R_{n,m} = 1$ if there exists an interaction between user $n$ and item $i$, and $R_{n,m} = 0$ otherwise. 
Based on the interaction data $\mathbf{R}$, a local graph $\mathcal{G}_n = (\mathcal{V}_n, \mathcal{E}_n)$ is constructed for each user $n$, where the set of nodes is the collection of the user $n$ and its interacted items $\mathcal{N}(u,n)$, namely $\mathcal{V}_n = \{ n \cup \mathcal{N}(u,n) \} $, and each edge represents an observed interaction between a user and an item $\mathcal{E}_n = \left\{ (n,m) : R_{n,m}=1 \right\}$. 
Then we use $\mathcal{G};\mathcal{V};\mathcal{E}$ to denote the union of all local graphs; the set of vertices of all local graphs; the set of edges of all local graphs respectively.
Using the set of local graphs $\{\mathcal{G}_n | n = 1,...,N\}$, we can train the recommender system to predict potential interactions for recommendation. Specifically, the goal of the system is to recommend the $K$ most relevant items to each user. 

We consider a federated recommendation system comprising a central server and $N$ distributed clients, where each client corresponds to a user and serves as a local device for storing historical interaction data. The central server acts as a central device for coordinating the training of a model across multiple clients. However, in contrast to traditional recommender systems, the central server cannot observe users' historical interactions due to privacy considerations. As a result, only recommendation models, rather than user data, can be exchanged between the server and user devices. Under such constraints, the system aims to train personalized recommendation models for individual clients. 
The essential symbols and notations utilized throughout this work are summarized in Table~\ref{tab:notation}.

\begin{table}
\caption{Table of Notations.}  
\begin{tabular}{l|l}  
\toprule 
\textbf{Notation} & \textbf{Definition} \\
\midrule
$\mathcal{U};\mathcal{I}$ & Set of users; Set of items. \\
$\mathcal{N}(u,n);\hat{\mathcal{N}}(u,n)$  & Set of interacted items of user $n$; Set of pseudo interacted items of user $n$.\\
$\mathcal{G};\mathcal{V};\mathcal{E}$ & The set of local graphs; Set of vertices; Set of edges. \\
$\mathbf{R}$ & Matrix of historical interactions. \\
$\mathbf{M}$ & The masking vector. \\
$L$ & The total number of layers in the graph neural network. \\
$K$ & The total number of clusters. \\
$N; M$ & The total number of clients (users); the total number of items. \\
$\alpha_{n,1};\alpha_{n,2};\alpha_{n,3}$ & The hyper-parameters controlling the personalization for client $n$.\\
$\mathbf{E}_{u,n};\mathbf{E}_{i,m}$ & The embedding for user $n$; the embedding for item $m$. \\
$\mathcal{L}_{rec}; \mathcal{L}_{bpr}; \mathcal{L}_{cl}$ & The recommendation loss; the BPR loss; the contrastive loss.\\
${g}_{n}; {g}_{n,e};{g}_{n,m}$ & The gradients for client $n$; the embedding gradients; the model gradients. \\
$\delta; \lambda$ & The parameters for local differential privacy. \\
\bottomrule 
\end{tabular}
\label{tab:notation}
\end{table}

\section{ Method }

\begin{figure}
    \centering
    \includegraphics[width=0.95\textwidth]{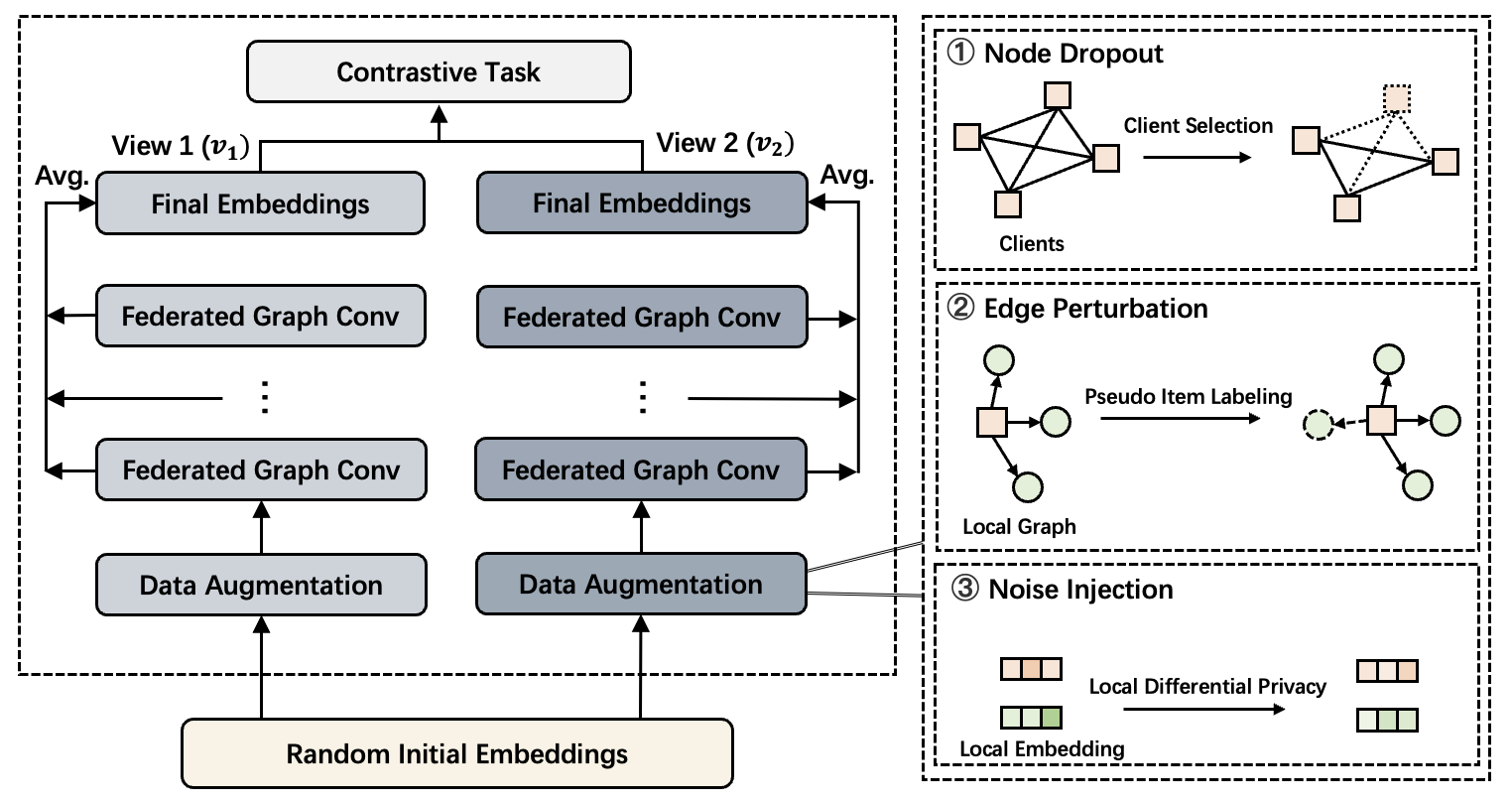}
    \caption{Self-Supervised Pre-Training Module.}
    \label{fig:sp}
\end{figure}

\begin{figure}
    \centering
    \includegraphics[width=0.95\textwidth]{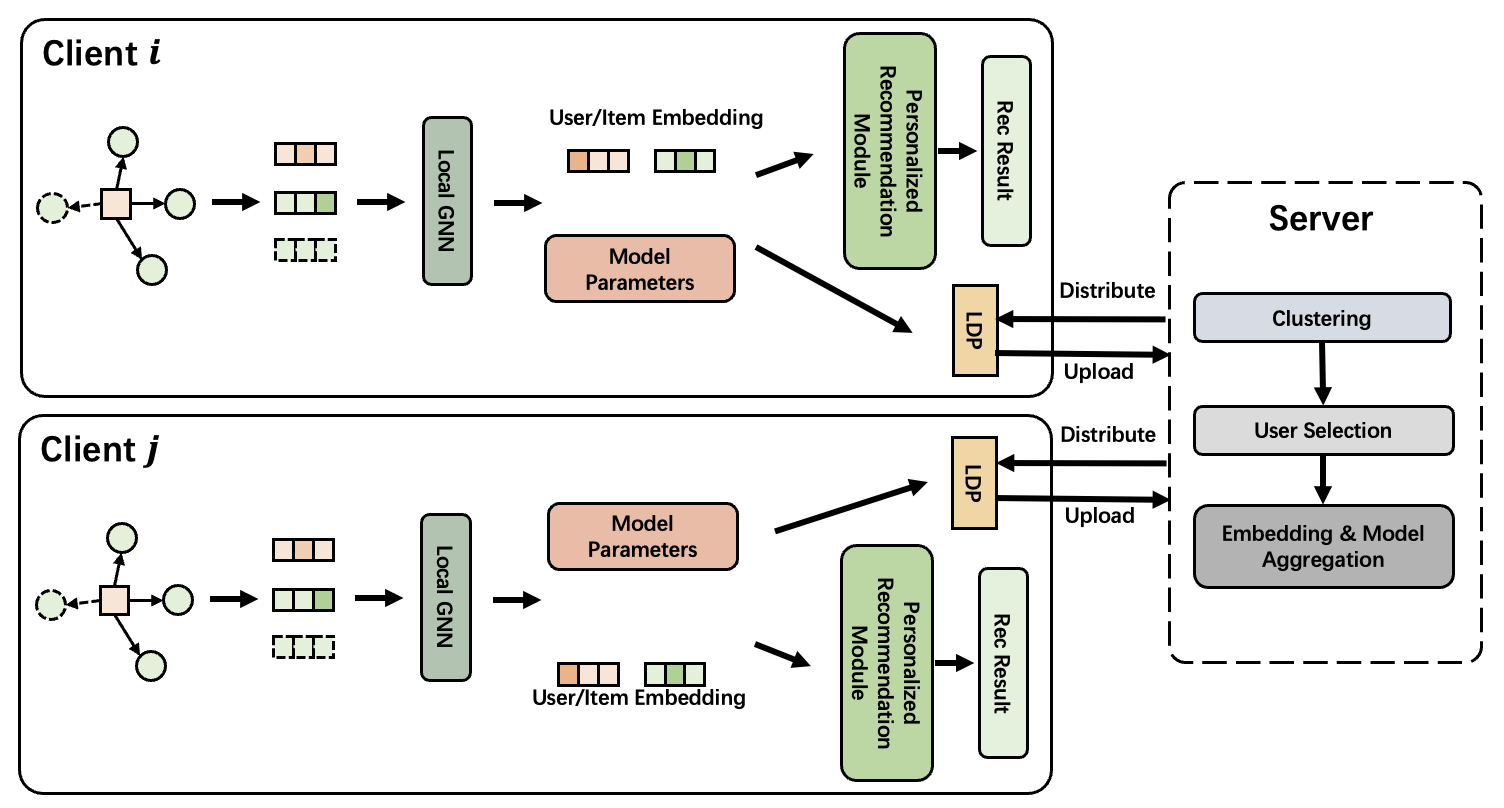}
    \caption{The training process of \proposed.}
    \label{fig:tp}
\end{figure}

We present the proposed \proposed paradigm, which leverages self-supervised pre-training to accelerate the training process and improve the model performance through more uniform representation learning. 
We first describe the self-supervised pre-training of \proposed, which entails constructing a contrastive signal from the input data (illustrated in Fig.~\ref{fig:sp}). Next, we elaborate on the federated training process of \proposed, comprising the user-side local recommendation network and server-side clustering-based federation (illustrated in in Fig.~\ref{fig:tp}). Finally, we summarize the proposed \proposed framework in Algorithm~~\ref{alg:fesog}.

\subsection{Self-Supervised Pre-Training Module}

There are two significant challenges in the current federated recommender systems. First, conventional federated recommender systems tend to converge slowly, resulting in a high communication burden between the server and clients, as observed in \cite{muhammad2020fedfast}. This issue can be particularly problematic in scenarios such as Internet of Vehicles, where the communication bandwidth is limited. Second, while privacy-preserving modules are effective in safeguarding user privacy, they have been found to adversely impact model performance. 

To mitigate these challenges simultaneously, we propose a training paradigm that employs a pre-training stage to expedite the training process and enhance model performance. Pre-training is a widely adopted technique in various domains, including computer vision (CV) \cite{bao2021beit,lu2019vilbert} and natural language processing (NLP) \cite{devlin2018bert,conneau2019cross}. In particular, we leverage self-supervised pre-training, wherein the server coordinates the clients to participate in the pre-training process. 
We first randomly initialize the user/item embeddings, and then use self-supervised pre-training to update the embeddings. The updated embeddings would be used as the initial embeddings for federated training in the latter step.

Next, we will describe the data augmentation process for self-supervised pre-training and formulate a contrastive loss for optimization. By utilizing this pre-training paradigm, we can address the aforementioned issues simultaneously, leading to improved model performance and faster convergence, thus reducing the communication burden between the server and clients.

\subsubsection{Data Augmentation for Self-Supervised Pre-Training.}
\label{sec:da4pt}
We have observed that the federated recommender system employs certain operations that facilitate the data augmentation process. For instance, in the federated training process, typically only a subset of clients participate, which is analogous to a data augmentation operation named \textit{node dropout} in self-supervised graph learning. We elaborate on the augmentation operators below.

\begin{itemize}[leftmargin=*]

\item \textbf{Node Dropout.}
In each round of the training process,  we randomly select a subset of clients to participate in the training. This is analogous to node dropout, where the clients that are not selected are dropped. This process can be formulated mathematically as follows:

\begin{equation}
\label{eq:aug1}
    \begin{split}
    & v_{1}(\mathcal{G}) = \left(\mathbf{M}^{\prime} \odot \mathcal{V}, \mathcal{E}\right), \\
    & v_{2}(\mathcal{G}) = \left(\mathbf{M}^{\prime \prime} \odot \mathcal{V}, \mathcal{E}\right),
    \end{split}
\end{equation}
where $\odot$ is the Hadamard product, and $\mathbf{M}^{\prime}, \mathbf{M}^{\prime \prime} \in \{0,1\}^{\left | \mathcal{V} \right |}$  are two masking vectors applied to the node set $\mathcal{V}$ to generate two subgraphs. This augmentation is expected to identify the influential nodes from differently augmented views and make the representation learning less sensitive to structural changes.

\item \textbf{Edge Perturbation.}
The pseudo item labeling technique is a widely adopted method for preserving privacy in federated recommendation systems, as described in Section~\ref{sec:localgnn}. We randomly select a subset of items and mask them as if they had interaction with the user. This technique is similar to the edge perturbation method employed in SGL \cite{wu2021self}, which randomly perturbs edges in a graph. Two independent processes are formulated as follows:
\begin{equation}
\label{eq:aug2}
    \begin{split}
    & v_{1}(\mathcal{G}) = \left(\mathcal{V},  \mathcal{E} \cup \mathcal{E}' \right), \\
    & v_{2}(\mathcal{G}) = \left(\mathcal{V},  \mathcal{E} \cup \mathcal{E}'' \right),
    \end{split}
\end{equation}
where $\mathcal{E}'$ and $\mathcal{E}''$ are randomly generated sets of edges. The new items introduced by $\mathcal{E}'$ and $\mathcal{E}''$ are called pseudo items. By perturbing edges, this technique can help identify important edges and make the representation learning more robust to changes in the graph structure.
\item \textbf{Noise Injection.}
Another commonly used privacy-preserving mechanism for federated recommender systems is local differential privacy, which is detailed in Section~\ref{sec:ldp}. In essence, we inject random noise into embeddings, which is similar to the data augmentation method used in SimGCL, where random noise is also added to the graph embedding \cite{yu2022graph2}.
The process is formulated as follows:

\begin{equation}
\label{eq:aug3}
    \begin{split}
    & E_{u,n}^\prime = E_{u,n} + \frac{\triangle_{u,n}^\prime}{\left\|  \triangle_{u,n}^\prime \right\|_2}, \\
    & E_{i,m}^\prime = E_{i,m} + \frac{\triangle_{i,m}^\prime}{\left\|  \triangle_{i,m}^\prime \right\|_2},
    \end{split}
\end{equation}
where $\triangle_{u,n}^\prime$ and $\triangle_{i,m}^\prime$ are random noise vectors added to the user and item embeddings $E_{u,n}$ and $E_{i,m}$, respectively.

\end{itemize}

We would mix these three augmentation methods to generate graph views since a single augmentation approach may not be enough. 

\subsubsection{Contrastive Loss}

We adopt the contrastive loss InfoNCE \cite{gutmann2010noise} for the self-supervised pre-training task. The idea is to maximize the agreement of positive pairs and minimize that of negative pairs. 
Suppose the $E'_u$ and $E''_u$ are the user embedding for two views and $E'_i$ and $E''_i $ are the item embedding for two views generated by adaptively edge drop, then the contrastive loss $\mathcal{L}_{cl}$ is denoted as:

\begin{equation}
\label{eq:cl_loss}
    \begin{split}
     &\mathcal{L}_{cl}^{user} = \sum_{u \in \mathcal{U}} -\text{log}\frac{\text{exp}(s(E'_u , E''_u/ \tau))}{  \sum_{u \in \mathcal{U}} \text{exp}(s(E'_u, E''_u)/ \tau)} ,\\
     &\mathcal{L}_{cl}^{item} = \sum_{i \in \mathcal{I}} -\text{log}\frac{\text{exp}(s(E'_i , E''_i/ \tau))}{  \sum_{i \in \mathcal{I}} \text{exp}(s(E'_i, E''_i)/ \tau)} ,\\
     &\mathcal{L}_{cl} =\mathcal{L}_{cl}^{user}+\mathcal{L}_{cl}^{item} ,\\
     \end{split}  
\end{equation}
where $s(\cdot)$ measures the similarity between two vectors, which is
set as cosine similarity function; $\tau$ is the hyper-parameter, known as
the \textit{temperature} in softmax. 
In this way, the representation learning could be enhanced and facilitate the model training. 
The self-supervised pre-training could make the federated training converge faster by providing a better initial state, thus could reduce the total number of epochs, and therefore reduce the communication overhead.

\subsection{User-Side Local Recommendation Network}
Our proposed user-side local recommendation network has three modules: a raw embedding module, a local GNN module, and a personalized prediction module.


\subsubsection{Embedding Layer}
The embedding layer is a crucial component in recommender systems as it is responsible for converting user/item ID information into dense embedding vectors \cite{luo2022hysage,wang2019neural,he2020lightgcn}. 
We leverage an embedding layer to transform user and item nodes into their respective embeddings, which we denote as $\mathbf{E}_u \in \mathbb{R}^{d \times N}$ and $\mathbf{E}_i \in \mathbb{R}^{d \times M}$, where $d$ is the embedding dimension.
Since we performed self-supervised pre-training in the previous section, here the embedding layer is initialized with the pre-trained parameters.

Following prior works \cite{liu2021federated}, item embeddings are shared and updated iteratively among users via the server, while the user embeddings are kept locally due to privacy concerns. These raw initial embeddings are used to train the local GNNs. During the training process, the global item embeddings and local user embeddings will be updated.

\subsubsection{Local GNN Module}
\label{sec:localgnn}
After getting its own embedding and the embeddings of all the items, each user needs the user-item interaction matrix to train the local GNN model. However, one difficulty is that user-item interaction information is kept private as local data and should not be shared among the server and other users.

In order to tackle such issues, we follow a similar idea as in \cite{wu2021fedgnn}, where each user uploads the privacy-protected embedding and the encrypted IDs (with the same encryption for all users) of the items that this user has interaction with to the server. Then the server sends encrypted item IDs and user embeddings back to all the users. To further reduce the communication cost, the server can just send back the encrypted item ID and the other users' embeddings to a user that has previously interacted with this item. Therefore, each user is able to get several users' embedding information together with the corresponding items, without revealing the identities of these users. In this way, each user could explore its neighborhood users and expand a local interaction graph.

If only the true interacted item embedding is uploaded, it can be easily inferred from the original user behavior. To address this issue, Wu \textit{et al.} \cite{wu2021fedgnn} proposed a pseudo item labeling technique that adds pseudo interacted item embeddings during the upload process to confuse the server. Furthermore, we propose a new technique, called \textit{interacted item masking}, to provide further protection to the system from potential attackers. The idea behind this technique is to randomly mask some real interacted items during the training process. These masked items are treated as non-interacted ones during local training. By doing so, this technique could further confuse potential attackers and enhance privacy protection.



The GNN module will output user $n$'s embedding $E_{u,n}$ and items' embeddings $\{E_{i,m}\}$: 

\begin{eqnarray}
\label{eq:localgnn}
     {E}^{(l)}_{\mathcal{N}(u,n)} &=& f_{aggregate}\left( \left\{{E}^{(l-1)}_v: v \in \mathcal{N} (u,n) \cup \hat{\mathcal{N}} (u,n) \right\}\right), \nonumber\\
    E_{u,n}^{(l)} &=& f_{combine}\left(E_{u,n}^{(l-1)}, {E}^{(l)}_{\mathcal{N}(u,n) }\right), \\
      E_{u,n} &=& f_{readout} \left( E_{u,n}^{(0)}, \ldots, E_{u,n}^{(L)} \right), \nonumber
\end{eqnarray}
where $E_{u,n}^{(l)}$ is embedding of user node $n$ at layer $l$, $\mathcal{N}(u,n)$ denotes the neighbor set of user $n$ in the local interaction graph $G$, 
 $\hat{\mathcal{N}}(u,n)$ denotes the pseudo neighbor set of user $n$, and $L$ is
the number of local GNN layers. The aggregate function $f_{aggregate}(\cdot)$ is applied to the representations of neighbors at the $(l-1)$-th layer to generate the representation of the $l$-th layer. 
This representation is then combined with the $(l-1)$-th layer's representation of user embedding using the $f_{combine}(\cdot)$ function. After $L$ iterations of propagation, the information of $L$-hop neighbors is encoded in $E_{u,n}$. 
Finally, the readout function $f_{readout}(\cdot)$ summarizes all the representations to obtain the final representation of user $n$ for recommendation. 
The representations of item nodes can be obtained in the same way.

Note that there are various choices for this plug and use GNN module, such as PinSage \cite{ying2018graph}, NGCF \cite{wang2019neural} and LightGCN \cite{he2020lightgcn}. 

These embeddings will be fed into the personalized prediction network for rating/preference predictions. 


\subsubsection{Personalized Prediction Module}
To achieve personalized recommendation, our framework trains personalized recommendation models for each user. Let us denote by $\Theta^{t}_{local,n}$ as the raw embedding and GNN model trained by user $n$ at time step $t$. From the server side, we will also have a global federated model $\Theta^{t}_{global}$ and a cluster-level federated model $\Theta^{t}_{C(n)}$, where $C(n)$ is the cluster containing user $n$. We will describe how to obtain the global and the cluster-level models later. The personalized model combines these three models together via weights $\alpha_{n,1},\alpha_{n,2},\alpha_{n,3}$, which can be either hyperparameters or learnable parameters:
\begin{equation}
    \begin{split}
     & \Theta^{t}_{n} = \alpha_{n,1} \Theta^{t}_{local,n} +\alpha_{n,2} \Theta^{t}_{C(n)}+ \alpha_{n,3} \Theta^{t}_{global}.
     \end{split}
\end{equation}

After obtaining the embeddings, an additional linear layer or a dot multiplication of the user and item embeddings could be used to get the rating prediction.

\subsubsection{Optimization}

After obtaining the user embedding and item embedding via GNN, we conduct the inner product, as a classical method, to estimate the user’s preference towards the target item.
Specifically, the model prediction for user $n$ towards item $m$ is denoted as:

\begin{equation}
    \begin{split}
     & \hat{y}_{n,m}= E_{u,n}^{\top} E_{i,m}.\\
     \end{split}
\end{equation}

We use the Bayesian personalized ranking (BPR) loss \cite{rendle2012bpr} to encourage the prediction of an observed user-item pair to be higher than its unobserved counterparts:

\begin{equation}
\begin{split}
& \mathcal{L}_{bpr} = - \sum_{n=1}^N \sum_{m \in \mathcal{N}_{u}} \sum_{j \notin \mathcal{N}_{u}} \text{ln} \sigma (\hat{y}_{n,m}-\hat{y}_{n,j}).
\end{split}
\label{eq:rec_loss}
\end{equation}

The overall recommendation loss is the sum of BPR loss and regularization loss, denoted as:
\begin{equation}
\begin{split}
\mathcal{L}_{rec} = \mathcal{L}_{bpr}  + \gamma \left \| \Theta \right \| ^2_2,
\end{split}
\label{eq:overall_loss}
\end{equation}
where $\Theta$ is the model parameters, and $\gamma$ is a hyperparameter controlling the strength of $L_2$ regularization loss.

\subsubsection{Local Differential Privacy}
\label{sec:ldp}

Local Differential Privacy (LDP) is a widely used and essential privacy-preserving technique in federated recommendation systems \cite{liu2021federated, wu2021fedgnn}. The LDP approach extends the concept of differential privacy (DP), which aims to obfuscate data by adding noise to the uploaded gradients \cite{dwork2008differential}. In our work, we employ the LDP module, which clips the local gradients if their $L_\infty$-norm is above a threshold $\delta$, and adds a zero-mean Laplacian noise to the unified gradients to achieve privacy protection. The gradient for client $n$ is denoted as $g_{n,e}$ and $g_{n,m}$ for the embedding and model gradients, respectively. For each gradient $g \in \{g_{n,e}, g_{n,m}\}$, we randomize it by applying the following formula:

\begin{equation}
\label{eq:ldp}
     {g}'=\operatorname{clip}\left(g, \delta\right)+\operatorname{Laplacian}\left(0, \lambda\right)
\end{equation}
where $g'$ is the randomized gradient, $\operatorname{clip}(\cdot, \delta)$ is the clipping function with the threshold $\delta$, and $\operatorname{Laplacian}\left(0, \lambda\right)$ is the Laplacian noise with zero mean and strength $\lambda$. As shown in \cite{wu2021fedgnn}, the privacy budget $\epsilon$ can be bounded by $\frac{2\delta}{\lambda}$. It is worth noting that a smaller $\delta$ and a larger $\lambda$ result in a smaller budget $\epsilon$, which indicates better privacy protection. Therefore, when selecting the values of $\delta$ and $\lambda$, we need to balance privacy protection and model performance carefully. 

By employing LDP, we can effectively protect the privacy of the user data while still achieving satisfactory model performance. The randomized gradients produced by LDP can help prevent malicious third parties from inferring sensitive information from the federated learning process.

\subsection{Server-Side Clustering Based Federation}
Our proposed server-side federation module performs three main functions: user clustering, user selection, and parameter aggregation. At each iteration, encrypted and privacy protected user/item embeddings and models are uploaded to the server by the users. 

\subsubsection{User Clustering}
Based on user embeddings $E_{u,n}$, the server clusters users into $K$ groups. 
User $n$ belongs to cluster $C(n)$. We can use any commonly used clustering method such as K-means ~\cite{macqueen1967some}. Since the node representation $E_{u,n}$ is jointly learned from the attribute and collaborative information at each user, the representation is therefore enhanced.

\subsubsection{User Selection}

In the context of federated recommender systems, traditional methods for client selection randomly choose clients from the entire pool \cite{wu2021fedgnn,liu2021federated}. However, in critical conditions where communication costs are high, our framework introduces an additional cluster-based user selection mechanism to facilitate training. Specifically, within each cluster, a few random users are adaptively selected in proportion to the cluster size to participate in the model aggregation in each iteration. This approach can improve the efficiency and effectiveness of federated recommender systems by selecting representative clients.

\subsubsection{Parameter Aggregation}
Our framework performs both network model aggregation and embedding aggregation. The user embedding is stored at the local user device but may get exchanged via the server without revealing the user identity. Item embeddings are shared and updated by all clients. For network models, the server will aggregate a global model $\Theta^{t}_{global}$ (via a weighted sum of all participating users) and cluster-wise models $\Theta^{t}_{C(n)}$ for cluster $C(n)$ (via a weighted sum of all participating users in the cluster). 
The gradient  $g_n^\prime$ of client $n$ is obtained in Section~\ref{sec:ldp}. We denote the local data amount in client $n$ as $D_n$, then the global model $\Theta^{t}_{global}$ is denoted as:
\begin{equation}
\label{eq:fedavg}
   \begin{split}
   & \overline{g} = \sum_{n \in \mathcal{S}} \frac{D_n}{\sum_{j \in \mathcal{S}} D_j} g_{n}' \\
& \Theta_{global}^t = \Theta_{global}^{t-1} - \eta \cdot \overline{g}
    \end{split}
\end{equation}
where $\eta$ is the learning rate and $\mathcal{S}$ is the set of selected clients.
The cluster-level model $\Theta^{t}_{C(n)}$ could be obtained in a similar way.
Then the global model $\Theta^{t}_{global}$ and the cluster-level model $\Theta^{t}_{C(n)}$ will be given to user $n$ for personalized recommendation.

\subsection{ Algorithm }
The pseudo-code of the algorithm of \proposed is presented in Algorithm~\ref{alg:fesog}. The input consists of the training hyper-parameters such as the embedding size $d$ and learning rate $\eta$. Additionally, the client data should also be given, \textit{i.e.} the client local graphs $\{\mathcal{G}_{n} | _{n=1}^{N}\}$. Though the target is to provide personalized recommendations, we output the parameters $\Theta$ and the local inferred embeddings $\{E_{u,n}:n=1,\ldots,N\}$, which is sufficient for clients to perform personalized recommendation. In the algorithm, line~\ref{alg:server_loop_start} to line~\ref{alg:server_loop_end} is the loop operated on the server, which sends parameters to clients and collects their gradients for updating. The function \textsf{ClientUpdate}() is the operation on local devices. It downloads the parameters to infer the local user embeddings (line~\ref{alg:downloading}). 
Pseudo-labeling and LDP are combined ~(line~\ref{alg:pesudo item sampling} and line~\ref{alg:LDP}) to protect the gradients from privacy leakage. 
This function returns the gradients and the number of interactions~(line~\ref{alg:func_ret}) for the server to collect. 


\begin{algorithm}[bth]
\caption{\textbf{\proposed} }
\label{alg:fesog}
\SetKwData{False}{False}\SetKwData{This}{this}\SetKwData{Up}{up}
\SetKwFunction{Union}{Union}
\SetKwComment{tcc}{/*\ }{}
\SetKwFunction{server}{ServerRun}
\SetKwFunction{ClientUpdate}{ClientUpdate}
\SetKwInOut{Input}{Input}\SetKwInOut{Output}{Output}
\SetKwProg{Fn}{Function}{:}{\KwRet}

\Input{Embedding size, learning rate: $d$, $\eta$ \\
Total number of clients, items: $N,M$ \\
The number of pseudo items: $p$ \\
LDP parameters: $\delta, \lambda$ \\
Client local graph: $\{\mathcal{G}_{n}: n=1,\ldots,N\}$
}

\Output{Model parameters and embeddings $\Theta$; Local client embeddings $\{E_{u,{n}}: n=1,\ldots,N\}$}
\BlankLine
 Initialize $\Theta$; \\
\tcc{Self-Supervised Pre-Training\  */}
 \For{$epoch \in \{1,...,k\}$}{
generate augmented view $v_1$ and $v_2$ $\leftarrow$ Eq.~(\ref{eq:aug1}), (\ref{eq:aug2}), (\ref{eq:aug3}) ;\tcp*[f]{data augmentation}\\
$\mathcal{L}_{cl}\leftarrow $ Eq.~(\ref{eq:cl_loss});\tcp*[f]{calculate contrastive loss}\\
compute gradient and update model;
 }
 \tcc{Federated Training\  */}

\While{not converge\label{alg:server_loop_start}}{
cluster users;
 sample a fractions of clients $\mathcal{S}$ based on clustering; \tcp*[f]{client selection}\\
\label{alg:clients_sampling}

\For{$n\in\mathcal{S}$}{
    ${g}_{n}$ = \ClientUpdate{$n$, $\Theta$}; \tcp*[f]{collect gradients from clients}\\
    }
$\bar{{g}}_{n}\leftarrow$ Eq.~(\ref{eq:fedavg}); \tcp*[f]{average gradients from clients} \\
$\Theta=\Theta - \eta\cdot\bar{{g}}_{n}$; \tcp*[f]{update parameters}\label{alg:server_loop_end}
}

\Fn{\ClientUpdate{$n$, $\Theta$}}{
download $\Theta$ from server;\label{alg:downloading} \\
${E}_{u,n} \leftarrow$ Eq.~(\ref{eq:localgnn});\label{alg:user_inference} \tcp*[f]{local GNN} \\
sample $p$ pseudo items;\label{alg:pesudo item sampling} \\
${\mathcal{L}}_{rec} \leftarrow $ Eq.~(\ref{eq:overall_loss}); \tcp*[f]{calculate recommendation loss}\\
${g}_{n} = \frac{\partial{\mathcal{L}}_{rec}}{\partial\Theta}$;  \tcp*[f]{calculate the gradients} \\
${{g}}_{n}'\leftarrow$ Eq.~(\ref{eq:ldp}); \tcp*[f]{incorporate LDP}\label{alg:LDP} \\
\KwRet ${{g}}_{n}'$;\label{alg:func_ret} \tcp*[f]{return gradients}\\ 
}
\end{algorithm}


\section{ Experiment }

In this section, we present the experimental evaluation of our proposed framework, \proposed, and its comparison with existing baseline methods. We demonstrate the effectiveness of our approach over real-world datasets. Furthermore, we conduct a comprehensive analysis of the framework's components to better understand their contributions.
Our research objectives are centered around the following research questions:

\begin{itemize}[leftmargin=*]
\item
\textbf{RQ1}: How does our model perform in comparison to baseline methods in terms of federated recommendation performance? 
\item
\textbf{RQ2}: How does the incorporation of graph contrastive learning and personalization module affect the federated recommendation performance?
\item
\textbf{RQ3}: How do hyper-parameters affect the model performance?
\end{itemize}

\subsection{ Experiment Setup }
\subsubsection{Datasets.}

To evaluate the effectiveness of our proposed framework, we conducted experiments on three real-world datasets, whose statistics are summarized in Table~\ref{dataset}. Yelp \cite{yelpdataset} is a widely-used benchmark dataset adopted from the 2018 edition of the Yelp challenge. Amazon-Kindle~\cite{he2016ups} was sourced from the Amazon review data and used to recommend e-books to users. Gowalla \cite{liang2016modeling} contains check-in data obtained from the Gowalla social network, where users share their locations by checking in.

\begin{table}[h]
\centering
\caption{Statistics of datasets.}
\label{dataset}
{
\begin{tabular}{c|cccc}
\toprule
Dataset   & \# of user & \# of item & \# of rating & density \\ \midrule
Yelp & 5,224      & 7,741      & 123,024      & 0.003042 \\ 
Amazon-Kindle & 7,650      & 9,173      & 137,124      & 0.001954  \\ 
Gowalla & 12,022      & 40,593      & 374,669      & 0.000768  \\ 
\bottomrule
\end{tabular}
}
\end{table}

\subsubsection{Baseline Methods.}

In order to illustrate the effectiveness of our proposed approach, we conduct a comparative analysis with several competitive baseline methods. To ensure a fair and unbiased comparison, we adopt LightGCN \cite{he2020lightgcn} as the backbone model for both our proposed approach and the baseline methods. The baseline methods can be classified into two categories: centralized methods and federated methods, depending on whether the data is stored locally. The centralized methods encompass traditional recommendation methods, the backbone model, and graph contrastive learning methods.
Below is detailed description of the baseline methods.

\noindent The traditional method: 
\begin{itemize}[leftmargin=*]
\item
\textbf{MF \cite{koren2009matrix}:} Matrix Factorization (MF) is a widely used approach that factorizes the user-item interaction matrix into low-rank matrices to predict user-item preferences.
\end{itemize}

\noindent The backbone model: 
\begin{itemize}[leftmargin=*]
\item
\textbf{LightGCN \cite{he2020lightgcn}:} it is a lightweight model which simplifies the weight parameter aggregation steps in graph convolution networks.
\end{itemize}

\noindent Graph contrastive learning methods: 
\begin{itemize}[leftmargin=*]
\item
\textbf{SGL \cite{wu2021self}:} incorporates the self-supervised learning into recommendation via graph level augmentation such as edge drop and random walk.
\item
\textbf{SimGCL \cite{yu2022graph2}:} injects the random noise to embeddings as an augmentation-free method for graph contrastive learning.
\item
\textbf{xSimGCL \cite{yu2022xsimgcl}:} is the state-of-the-art graph contrastive learning for recommendation method. It is the simplified and more powerful version of SimGCL.
\end{itemize}

\noindent Federated recommendation methods:
\begin{itemize}[leftmargin=*]
\item
\textbf{FedMF \cite{chai2020secure}:}
it utilizes a secure aggregation protocol
to enable multiple parties to collaboratively train a MF model on their private data without sharing it.
\item
\textbf{FedGNN \cite{wu2021fedgnn}:}
proposes a federated framework for privacy-preserving GNN-based recommendation by locally training GNN models while applying local differential privacy techniques to protect user privacy.
\item
\textbf{PerFedRec \cite{luo2022personalized}:}
is the state-of-the-art federated recommendation method. 
The clustering technique is incorporated as well as joint representation learning 
to make personalized federated recommendations.
\end{itemize}



\subsubsection{Implementation and Training Settings.}

In our experiments, we use a lightweight model named LightGCN \cite{he2020lightgcn} as the GNN model, and use the dot product to implement the rating predictor. The user and item embeddings and their hidden representations learned by graph neural networks are 64-dimensional. 
Following previous works \cite{deshpande2004item}, we apply the leave-one-out strategy for evaluation, and employ Recall@K and NDCG@K to evaluate the performance. For each user, we use the last behavior for testing, the second to last for validation and the others for training. 
The number of users used in each round of model training is 256, and the default learning rate is 0.01. The hyper-parameters $\alpha_{n,1}$, $\alpha_{n,2}$, $\alpha_{n,3}$ are set to $\frac{1}{3}$.
The hyper-parameters (\textit{e.g.}, learning rate, dropout rate) of baseline methods are selected according to the best performance
on the validation set. 
The performance is averaged over five runs on the testing set.
We used PyTorch \cite{paszke2019pytorch} for the implementation of all experiments.

\begin{table}[htb]
\caption{Experiment results compared with baseline methods. The best federated learning results are in bold, and the best results for non-federated learning methods are underlined. Improvement indicates the percent that \proposed improves against the second-best federated learning result. All improvements are significant with $p$-value < 0.05.}
\label{tab:main_exp}
\resizebox{\textwidth}{!}{
\begin{tabular}{c|cccccc}
\toprule
\multirow{2}{*}{Method} & \multicolumn{2}{c}{Yelp} & \multicolumn{2}{c}{Amazon-Kindle}  & \multicolumn{2}{c}{Gowalla} \\
\cmidrule(r){2-3} \cmidrule(r){4-5} \cmidrule(r){6-7}
&  Recall      &  NDCG   
&  Recall      &  NDCG
&  Recall      &  NDCG   \\
\midrule
MF & $0.0546_{\pm0.0038}$&$0.0195_{\pm0.0013}$&$0.1077_{\pm0.0011}$&$0.0442_{\pm0.0012}$  & $0.1211_{\pm0.0037}$&$0.0528_{\pm0.0027}$  \\
LightGCN &$0.0690_{\pm0.0018}$&$0.0265_{\pm0.0010}$& $0.1196_{\pm0.0029}$  & $0.0499_{\pm0.0006}$  & $0.1493_{\pm0.0015}$&$0.0699_{\pm0.0013}$  \\
SGL & $0.0752_{\pm0.0021}$&$0.0289_{\pm0.0008}$&$0.1265_{\pm0.0033}$&$0.0523_{\pm0.0016}$  & $0.1549_{\pm0.0027}$&$0.0728_{\pm0.0027}$ \\
SimGCL & $\underline{0.0779}_{\pm0.0013}$&$\underline{0.0297}_{\pm0.0008}$&$0.1277_{\pm0.0015}$&$0.0536_{\pm0.0003}$  & $0.1592_{\pm0.0003}$&$0.0757_{\pm0.0007}$  \\
xSimGCL &$0.0767_{\pm0.0018}$&$0.0289_{\pm0.0008}$& {$\underline{0.1289}_{\pm0.0013}$}&{$\underline{0.0540}_{\pm0.0008}$}  & {$\underline{0.1620}_{\pm0.0018}$}&{$\underline{0.0775}_{\pm0.0013}$}  \\
\midrule
FedMF &$0.0441_{\pm0.0027}$&$0.0169_{\pm0.0005}$&$0.0856_{\pm0.0026}$&$0.0358_{\pm0.0013}$&$0.1101_{\pm0.0029}$&$0.0525_{\pm0.002}$  \\
FedGNN &$0.0455_{\pm0.0025}$&$0.0173_{\pm0.001}$&$0.0898_{\pm0.0012}$&$0.0381_{\pm0.0007}$&$0.1111_{\pm0.0011}$&$0.0552_{\pm0.0008}$  \\
PerFedRec & $0.0453_{\pm0.0012}$&$0.0171_{\pm0.0006}$&$0.0905_{\pm0.0016}$&$0.0393_{\pm0.0009}$&$0.1126_{\pm0.0005}$&$0.0554_{\pm0.0006}$ \\
\proposed & $\textbf{0.0534}_{\pm0.0017}$&$\textbf{0.0201}_{\pm0.0014}$&$\textbf{0.1013}_{\pm0.0033}$&$\textbf{0.0443}_{\pm0.0017}$&$\textbf{0.1199}_{\pm0.0016}$&$\textbf{0.0588}_{\pm0.0005}$ \\

\midrule
Improvement & \text{17.36\%} & \text{16.18\%} & \text{11.93\%} & \text{12.72\%} & \text{6.48\%} & \text{6.14\%}\\

\bottomrule
\end{tabular}}
\end{table}

\subsubsection{Hardware Descriptions.}
Our experiments were conducted on a high-performance server running Ubuntu 18.04 LTS and equipped with 128 GB of RAM. The server was powered by six Intel(R) Xeon(R) Platinum 8268 CPUs with a clock speed of 2.90GHz, and featured two NVIDIA Tesla V100-SXM2-32GB GPUs for accelerated computing.

\subsection{ Performance Comparison }

In this section, we present an overall comparison of different models with experimental results shown in Table~\ref{tab:main_exp}. The results are categorized into two groups of centralized and localized methods. The following observations can be made from the experimental results:

\begin{itemize}
    \item GNN-based methods (\textit{e.g.}, LightGCN) outperform non-GNN methods (\textit{e.g.}, MF). This is because GNN-based models can directly model structure information and learn high-order collaborative information. Moreover, between the two federated learning baselines, FedGNN significantly outperforms FedMF, which further supports the claim that GNN-based models are superior to MF-based approaches.
    
    \item Our proposed method significantly outperforms the state-of-the-art federated recommender systems in all datasets. Compared with PerFedRec, \proposed achieves an average of $11.92\%$ and $11.68\%$ relative improvements in Recall and NDCG, respectively. Several advantages of \proposed demonstrate its superiority, including the importance of uniform representation learning through contrastive signals and personalized recommendations. 
    
    \item In almost all scenarios, the centralized method achieves the best results. Federated learning, on the other hand, impairs the performance compared with centralized learning. There are two reasons for this. Firstly, to achieve privacy protection, the federated learning framework has no access to the local data, limiting its capacity to model global structures. Secondly, the local gradients are protected by adding random noise, which prevents the server from receiving qualitative gradients from clients, even though it theoretically does not hurt the performance in expectation. This observation also highlights the importance of finding a trade-off between model performance and privacy protection.
    We partially address this issue by combining the noise with the contrastive learning method.
\end{itemize}

\begin{table}[htb]
\caption{Ablation study on Gowalla dataset. The best results are in bold.}
\label{tab:abl}
\begin{tabular}{c|cccc}
\toprule
\multirow{2}{*}{Variants} & \multicolumn{2}{c}{Recall} & \multicolumn{2}{c}{NDCG}   \\
\cmidrule(r){2-3} \cmidrule(r){4-5} 
 &  @10      &  @20   
 &  @10      &  @20
   \\
\midrule
\proposed & $\textbf{0.0862}_{\pm0.0010}$&$\textbf{0.1199}_{\pm0.0016}$&$\textbf{0.0503}_{\pm0.0005}$&$\textbf{0.0588}_{\pm0.0005}$  \\
\textit{w/o} Self-Supervised Pre-Training & $0.0816_{\pm0.0010}$&$0.1126_{\pm0.0005}$&$0.0476_{\pm0.0008}$&$0.0554_{\pm0.0006}$ \\
\textit{w/o} Personalized Recommendation & $0.0854_{\pm0.0007}$&$0.1190_{\pm0.0012}$&$0.0502_{\pm0.0004}$&$\textbf{0.0588}_{\pm0.0003}$ \\
\textit{w/o} Clustering & $0.0860_{\pm0.0014}$&$0.1197_{\pm0.0013}$&$0.0502_{\pm0.0009}$&$0.0586_{\pm0.0006}$  \\
\bottomrule
\end{tabular}
\end{table}
\subsection{Ablation Study}
We conduct an ablation study to evaluate the contribution of each module in our proposed framework to the overall performance. The results are presented in Table~\ref{tab:abl}. It is observed that the full model outperforms all the variants on three splits, indicating that all the main components contribute to performance improvement. Further description and analysis of each component are presented below.
\begin{itemize}[leftmargin=*]
\item \textbf{\textit{w/o} Self-Supervised Pre-Training}: In this variant, we remove the self-supervised pre-training module and only maintain the federated training process. 
The evaluation results show a significant degradation in the model performance without self-supervised pre-training. 
This finding suggests that self-supervised pre-training can effectively enhance the representation of the data and learn more uniform embeddings, which in turn improves the federated recommendation performance.

\item \textbf{\textit{w/o} Personalized Recommendation Module}: In this variant, we remove the personalized recommendation module and utilize the same recommendation model for all clients. The evaluation results reveal that the model performance without the personalized recommendation module decrease. This observation validates the effectiveness of the personalized recommendation module, which adapts the model to the specific preferences of each client and improves the overall performance.

\item \textbf{\textit{w/o} Clustering}: In this variant, we replace the user clustering with random choice, while keeping the other components the same. 
The evaluation results indicate that the proposed framework without user clustering performs relatively poorly. 
This finding highlights the importance of user clustering in enhancing the model's performance. User clustering allows the model to group similar users, which improves the model's ability to learn from the clients' data and make personalized recommendations.

\end{itemize}


\begin{figure}[htb]
 \subfigure[Yelp]{
\includegraphics[width=.32\textwidth]{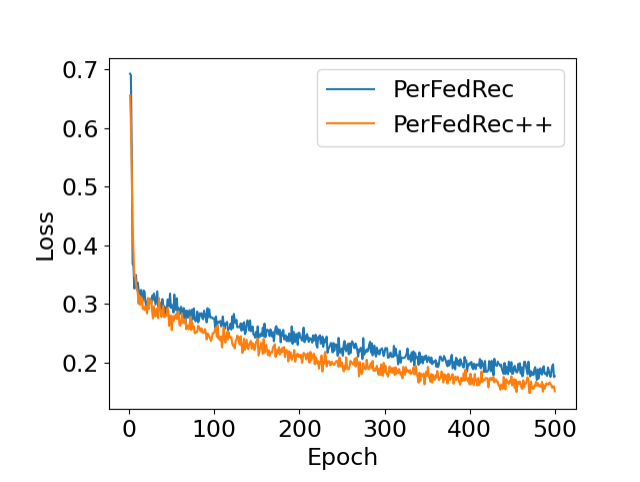}}
 \subfigure[Amazon-Kindle]{
\includegraphics[width=.32\textwidth]{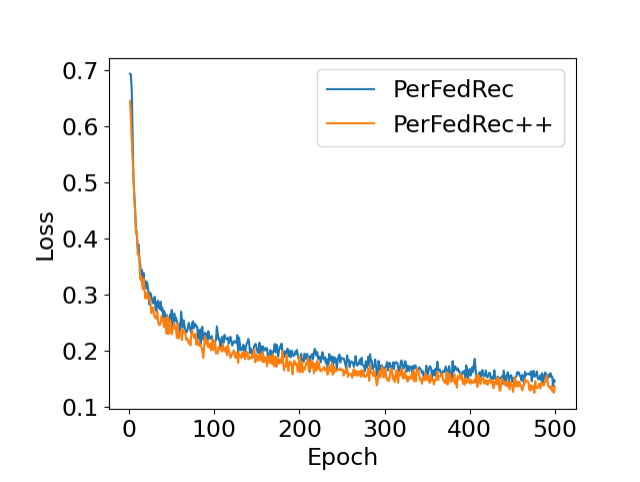}}
 \subfigure[Gowalla]{
\includegraphics[width=.32\textwidth]{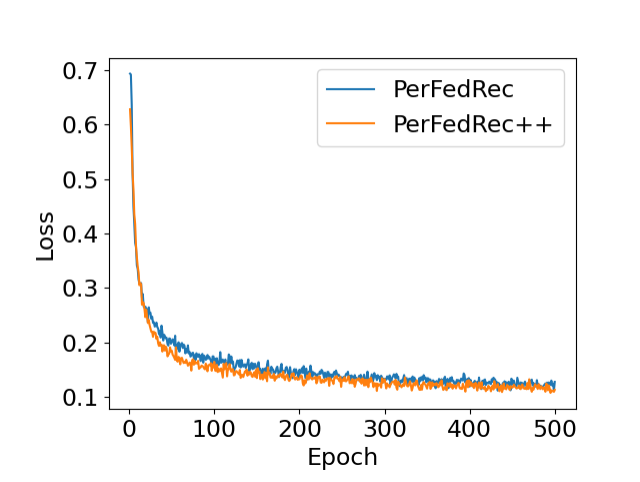}}
\caption{Loss Curve \textit{w.r.t} training epoch on three datasets.}
\label{fig:loss_curve}
\end{figure}

\begin{figure}[htb]
 \subfigure[Yelp]{
\includegraphics[width=.32\textwidth]{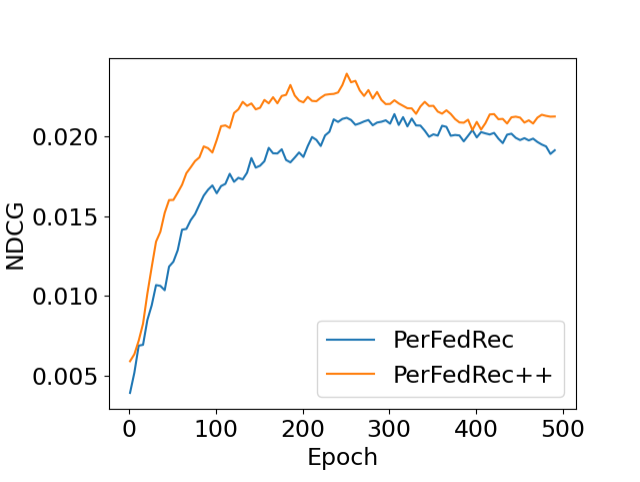}}
 \subfigure[Amazon-Kindle]{
\includegraphics[width=.32\textwidth]{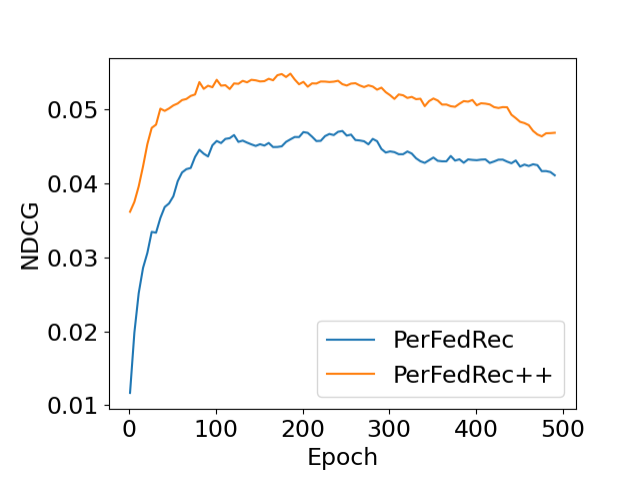}}
 \subfigure[Gowalla]{
\includegraphics[width=.32\textwidth]{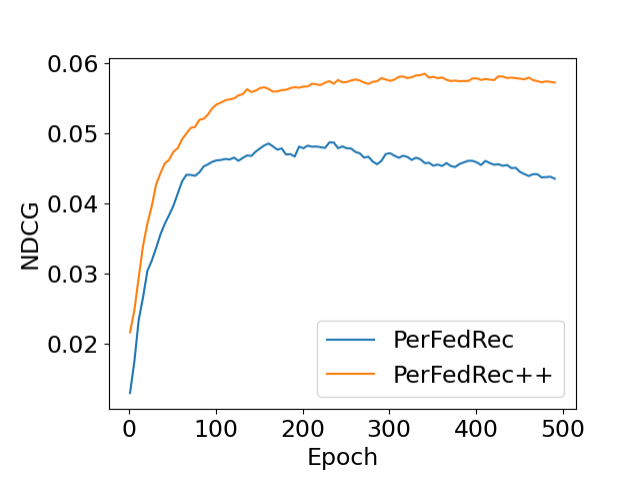}}
\caption{Validation NDCG@20 Curve \textit{w.r.t} training epoch on three datasets.}
\label{fig:ndcg_curve}
\end{figure}

\subsection{ Analysis of Training Efficiency}

We would like to study the training efficiency of \proposed. 
Figure~\ref{fig:loss_curve} and Figure~\ref{fig:ndcg_curve} show the training curves of \proposed and PerFedRec on three datasets. 
We have the following observations:

\begin{itemize}
\item Clearly, \proposed exhibits much faster convergence than PerFedRec on the Yelp and Amazon-Kindle datasets. Specifically, \proposed attains the best performance at approximately 250 and 200 epochs, respectively, while PerFedRec requires roughly 300 and 250 epochs to reach the best performance levels. This finding indicates that our proposed approach can significantly reduce the training time while still achieving remarkable improvements in performance. We ascribe this speedup to the use of a contrastive loss as the pre-training objective, which enables the model to learn representations from multiple negative samples. In contrast, the BPR loss in recommendation only uses one negative sample, which limits the model's perception field.

\item Another key observation is that \proposed outperforms PerFedRec, resulting in lower loss and higher validation NDCG scores. This improvement can be attributed to the self-supervised pre-training module, which learns more uniform representations. By learning from multiple negative samples, the self-supervised pre-training enhances the model's ability to capture the underlying patterns in the data, resulting in improved performance in the federated recommendation task.
\end{itemize}






\subsection{ Hyper-parameter Study }

We performed a comprehensive hyper-parameter study to investigate the impact of various hyper-parameters on the performance of the \proposed. We varied the following hyper-parameters and evaluated their impact on the system performance.

\subsubsection{Impact of Embedding Size}

\begin{figure}[htb]
 \subfigure[Yelp]{
\includegraphics[width=.32\textwidth]{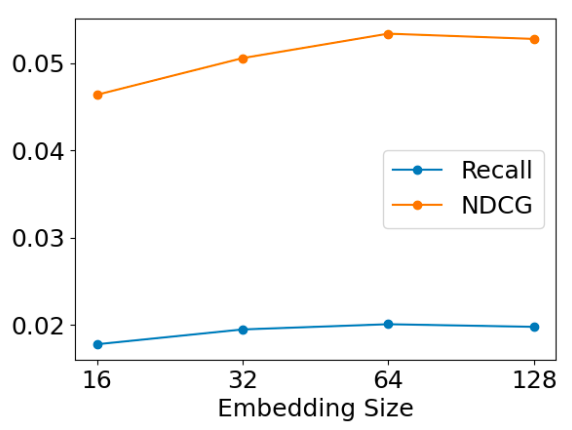}}
 \subfigure[Amazon-Kindle]{
\includegraphics[width=.32\textwidth]{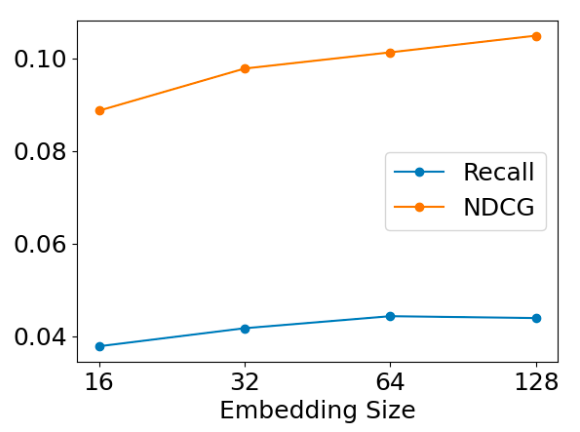}}
 \subfigure[Gowalla]{
\includegraphics[width=.32\textwidth]{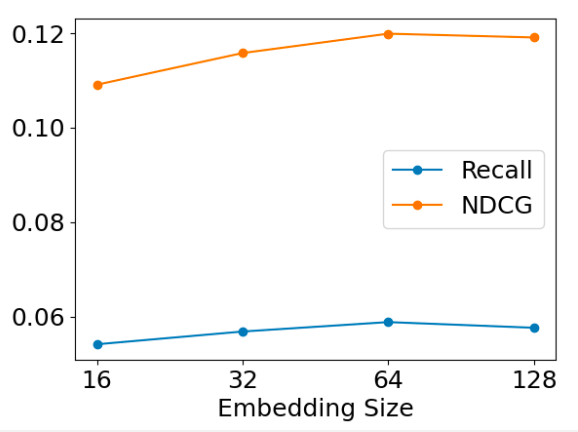}}
\caption{Impact of embedding size on model performance on three datasets.}
\label{fig:emb_size}
\end{figure}

The impact of embedding size is a crucial consideration in designing a federated recommender system as it determines the dimensionality of the latent space where the user and item representations are learned. In this section, we study the correlation between model performance and varying embedding size. The results are shown in Figure~\ref{fig:emb_size}. We observe that the embedding size can have a significant impact on the performance of the system. A larger embedding size can provide a higher capacity for the model to capture complex patterns in the data, but it can also lead to higher communication cost. On the other hand, a smaller embedding size can limit the expressive power of the model, but it could reduce the communication burden. Therefore, the optimal embedding size should be considered comprehensively in practice.

\subsubsection{Impact of Pre-Training Epoch}

\begin{figure}[htb]
 \subfigure[Yelp]{
\includegraphics[width=.32\textwidth]{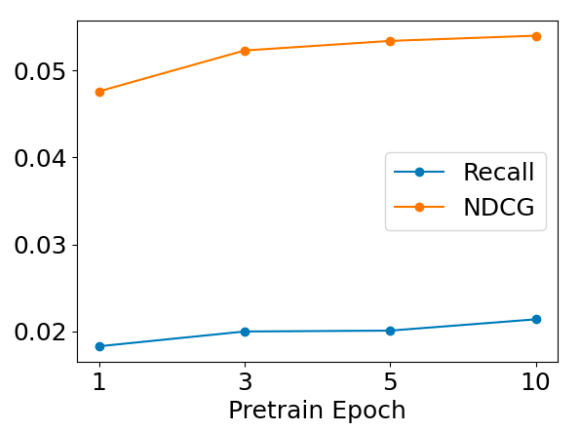}}
 \subfigure[Amazon-Kindle]{
\includegraphics[width=.32\textwidth]{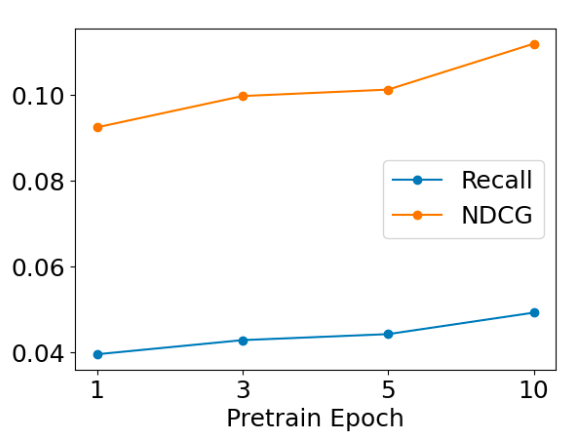}}
 \subfigure[Gowalla]{
\includegraphics[width=.32\textwidth]{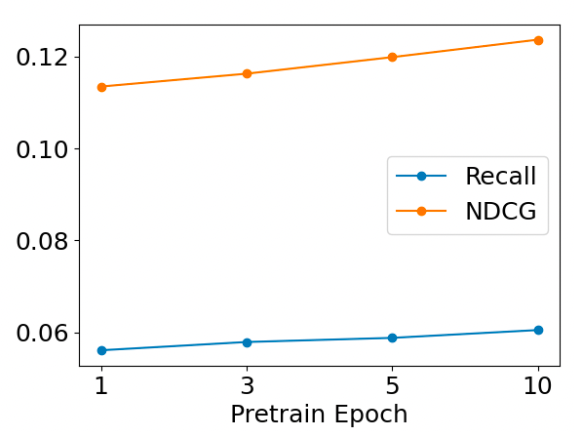}}
\caption{Impact of pre-training epoch on model performance on three datasets.}
\label{fig:pretraining_epoch}
\end{figure}

In \proposed, pre-training is a crucial step for learning user and item representations before fine-tuning the model on the local data of each participant. The pre-training process aims to initialize the model parameters with a good starting point to facilitate faster and more effective convergence during the fine-tuning stage.
In this section, we investigate the impact of the number of pre-training epochs on the performance of the federated recommender system. Specifically, we vary the number of pre-training epochs in the range of $\{1,3,5,10\}$ and evaluate the system's performance using appropriate metrics. The results are shown in Figure~\ref{fig:pretraining_epoch}.
We observe that a larger number of pre-training epochs generally leads to better performance. This is because the model has more opportunities to learn from the training data and capture the underlying patterns in the data more effectively. However, it is important to note that increasing the number of pre-training epochs could also lead to high communication cost and overfitting and may not necessarily lead to better performance beyond a certain point.
On the other hand, a smaller number of pre-training epochs could limit the model's ability to learn the underlying patterns in the data, resulting in suboptimal performance. In practice, the optimal number of pre-training epochs should be determined based on the available data, the complexity of the model, and the desired level of performance.

\section{Conclusion}
In this paper, we highlight the importance of self-supervised pre-training in personalized federated recommendations and propose a novel personalized federated recommendation framework named \proposed. 
We propose a self-supervised pre-training strategy to learn more uniform representations, which could enhance model performance and alleviate communication burdens.
To address the challenge of heterogeneity,  the proposed framework jointly learns user representations from collaborative and attribute information via GNNs, clusters similar users, and obtains personalized recommendation models by combining the user-level, cluster-level, and global models.
Experiments on three real-world datasets demonstrate that our proposed framework achieves superior performance for federated recommendation.


\bibliographystyle{ACM-Reference-Format}
\bibliography{ref}
\end{document}